\documentstyle[aps,epsf]{revtex}
\begin{document}

\title{Implications of $su(2)$ symmetry 
on the dynamics of population difference in the 
two-component atomic vapor}
\author{A. B. Kuklov$^1$ and Joseph L. Birman $^2$}
\address{$^1$ Department of Engineering 
Science and Physics,
The College of  Staten Island, CUNY,
     Staten Island, NY 10314}
\address{$^2$ Department of Physics, City College, CUNY, New York, NY 10031}

\date{\today}
\maketitle

\begin{abstract}
We present an exact many body solution
for the dynamics of the population difference ~$N_2-N_1$~
induced by an rf-field 
in the two-component atomic cloud characterized
by equal scattering lengths. 
We show that no intrinsic decoherence 
occurs for ~$N_2-N_1$~, 
provided the exact intrinsic $su(2)$ symmetry holds.
Decoherence for ~$N_2-N_1$~
arises when deviations from the symmetry exist
either in the scattering lengths or in the confining
potentials. 
We suggest testing the evolution of ~$N_2-N_1$~ 
by conducting echo-type experiments.
\\


\end{abstract}\vskip0.5 cm

Successful achievement of Bose-Einstein condensation
in  atomic traps
\cite{BEC} stimulated strong interest
in the
mechanisms of dissipation and decoherence
in the confined atomic clouds (see in \cite{REV}).
Many intriguing questions
are associated with the finiteness of
the number of bosons $N$ forming the 
Bose-Einstein condensate (BEC) in the trap.
It has been realized that the global phase
may experience collapses and revivals as long as
$N$ is finite \cite{SOL,WALLS,YOU,MILBURN,JUHA}.

The experimental study of the temporal 
correlations of the relative phase of the two-component
~$^{87}$Rb BEC has been conducted by
the JILA group \cite{JILA} by means of measuring 
the dynamics of the population
difference ~$N_2-N_1$~ between the 
components.
No
decay of the global relative 
phase has been detected on the time scale
of the experiment $\leq 100$ms. 
The question was then posed in \cite{JILA} 
as to why the phase correlations are
so robust despite an apparent fast relaxation of
other degrees of freedom such as, e.g.,
the relative motion of the two condensates.

In the present work, we address the question
\cite{JILA} of the relative phase correlations.
We show that if the two-component system
obeys the intrinsic $su(2)$ symmetry, as
is almost the case for $^{87}$Rb \cite{$su(2)$},
no intrinsic decoherence exists for the 
operators which are the generators of this symmetry.
In particular, the population
difference falls into 
this category. This result is general and does not
depend on a particular form of the Hamiltonian.
On the contrary, the nature of the
decoherence arising due to
deviations from the symmetry is not universal, and 
it depends on details of the Hamiltonian.
We discuss  how the nature of the decoherence
of ~$N_2-N_1$~ can be tested in the echo-type experiment
(see, e.g., in \cite{ECHO}). 
Specifically, we suggest modifying
the JILA experiment \cite{JILA} by introducing
an extra pulse between the
two $\pi /2$ pulses employed in \cite{JILA}. 
The echo should then be seen by the 
readout pulse (the last one) 
at the time $t_e=2\tau $ where $\tau$ is the time
delay between the first two pulses.  We note that
the atomic echo in optical lattices, predicted by one
of the present authors \cite{ECKUK},
has been observed very recently \cite{ECEXP}
under, however, the conditions when the inter-particle
interaction is not important.

We consider the two-component 
bosonic system with no exchange interaction
between the components (no intrinsic inter-conversion
between the species). 
This system can be described by the
second quantized Hamiltonian $H$ \cite{HAM}.
We note that this Hamiltonian
has been employed
for studying the dynamics of ~$N_2-N_1$~
in ~$^{87}$Rb in the mean
field approximation (MF) \cite{JILA2,WILLI}.
In what follows we will obtain the class of {\it exact solutions}
of the full many-body quantum mechanical problem in the
case of the $su(2)$ symmetry. 
We represent ~$H$~ \cite{HAM} as

\begin{eqnarray}
\displaystyle
H&=&H_S + H_A +\epsilon_0J_z + H_P,
\label{1_1}
\\
H_S&=&\int d{\bf x} \{\Psi^{\dagger}_1H_{1s}
 \Psi_1
+\Psi^{\dagger}_2H_{1s} \Psi_2 + 
{g\over 2}(\Psi^{\dagger}_1\Psi_1
+\Psi^{\dagger}_2\Psi_2)^2 +
{\tilde{g}\over 2}(\Psi^{\dagger}_1\Psi_1
-\Psi^{\dagger}_2\Psi_2)^2\}, 
\label{1_2}
\\ 
H_A&=&\int\, d{\bf x} \,\{ 
\Psi^{\dagger}_1H_{1a} \Psi_1
-\Psi^{\dagger}_2H_{1a} \Psi_2 +
{g'\over 2}[(\Psi^{\dagger}_1\Psi_1)^2
-(\Psi^{\dagger}_2\Psi_2)^2]\} ,
\label{1_3}
\\
H_P&=&{\hbar \Omega^* (t)\over 2} J_+ +{\hbar \Omega (t)\over 2} J_-\,\, .
\label{1_4}
\end{eqnarray}
\noindent
Here
 ~$ \Psi_{i}$~ is the second quantized
Bose field of the $i$-the component ($i=1,2$);
~$H_S$ is symmetric and $H_A$~ is anti-symmetric
with respect to exchanging the components ~$\Psi_{1,2}\to
\Psi_{2,1}$; ~$ H_P$~
describes the effect of the rf-field \cite{JILA,JILA2,WILLI};
~$\epsilon_0=const$~
is the detuning of the external field ~$\hbar\Omega (t)$~
(taken in the rotating wave approximation ) from 
resonance between the components;
~$ \Omega(t)$~ stands for the 
corresponding Rabi frequency
treated as an envelope of the rf-pulses \cite{JILA,JILA2};
~$H_{1s}=(H_1+H_2)/2,\,\, H_{1a}=(H_1-H_2)/2$~;
~$H_j=-\hbar^2\nabla^2/2m \,\, + U_j({\bf x})$~ ($j=1,2$),
stands for the one-particle part which
includes the kinetic energy and the trapping
potential; 
the binary collision terms
are taken in the contact form,
where

\begin{eqnarray}
g={\pi \hbar^2 \over m}(a_1+a_2 +2a_{12}),\,\,\,\,
\tilde{g}={\pi \hbar^2 \over m}(a_1+a_2 -2a_{12}),\,\,\,\,
g'={2\pi \hbar^2 \over m}(a_1-a_2)
\label{2}  
\end{eqnarray}
\noindent
are related to the corresponding scattering lengths
~$a_1,\,a_2,\,
a_{12}$.
We have also introduced the following operators

\begin{eqnarray}
J_z={1\over 2}\int d{\bf x}(\Psi^{\dagger}_2\Psi_2-
\Psi^{\dagger}_1\Psi_1);\,\,\,
J_+=\int d{\bf x}\Psi^{\dagger}_2\Psi_1 ; \,\,\,
J_-=\int d{\bf x}\Psi^{\dagger}_1\Psi_2 \,.
\label{3}
\end{eqnarray}
\noindent
It is important that
the operators (\ref{3}) represent the $su(2)$ algebra
of the angular momentum operators.
Indeed, it is easy to see that these operators obey the
standard $su(2)$ commutation relations 

\begin{eqnarray}
[J_z,J_+]=J_+,\,\,\,\,
[J_z,J_-]=-J_-,\,\,\,\, [J_+,J_-]=2J_z,
\label{3_1}
\end{eqnarray}
\noindent
provided the 
field operators obey the Bose commutation rule
~$[\Psi_i({\bf x}),\Psi^{\dagger}_j({\bf x}')]=\delta_{ij}
\delta ({\bf x}-{\bf x}')$. 
The operator ~$J_z=(N_2-N_1)/2 $,
with ~$N_{1,2}$~ standing for the total number
of bosons in the first and the second components,
respectively,
can be viewed
as the z-component of the angular momentum operator;
~$J_x=(J_++J_-)/2,\,\,J_y=(J_+-J_-)/2i$~
are the x,y-components, respectively. We note that
the formalism of the effective angular momentum 
operators has been successfully applied in many 
areas including optics (see, e.g., in Refs.
\cite{MILBURN,JS}). 

In the work \cite{$su(2)$} it has been 
emphasized that the two component ~$^{87}$Rb
realizes the $su(2)$ intrinsic symmetry to a very good
approximation due
to almost perfectly satisfying the relations
 ~$a_1= a_2=a_{12}$ as well as 
~$U_1=U_2$.  Thus, 
~$\tilde{g}=g'=0$~ and ~$H_{1a}=0$~ in
Eqs.(\ref{2}), (\ref{1_2}),
(\ref{1_3}). 
Then, the part ~$H_A$~ (\ref{1_3})
is zero, and in the case ~$\epsilon_0=0,\,\, H_P=0$~,
the total Hamiltonian is just ~$H_S$~ (\ref{1_2}) which
commutes with all the $su(2)$ operators
(\ref{3}) 
 even though ~$g\neq 0$.
This situation corresponds to an exact $su(2)$
symmetry \cite{$su(2)$} in the trap \cite{COM1}. 

Let us, first, consider the effect of 
violation of the $su(2)$ symmetry by the
external rf-field (no intrinsic violation).
So we choose ~$g'=0,\,\,\tilde{g}=0,\,\, H_{1a}=0$~
in the Hamiltonian (\ref{1_1})-(\ref{1_4}).
It is important to emphasize that in this case,
the full Hamiltonian
(\ref{1_1}) forms a closed algebra with
the operators (\ref{3}). Accordingly,
the Heisenberg equations for 
 ~$J_{+,-,z}$~
are 

\begin{eqnarray}
i\hbar \dot{J}_z={\hbar \Omega^* (t) \over 2}J_+ -
{\hbar \Omega (t) \over 2}J_-,\,\,\,
\,\, i\hbar \dot{J}_+=-\epsilon_0J_+
+\hbar\Omega(t)J_z,\,\,\,
i\hbar \dot{J}_-=\epsilon_0J_-
-\hbar\Omega^*(t)J_z\, ,
\label{4}
\end{eqnarray}
\noindent
where the $su(2)$ commutation relations (\ref{3_1})
have been
employed. We note that these equations
are exact and linear
in ~$J_{+,-,z}$~. Accordingly, 
there is no dissipation for these quantities ( unless
the parameters ~$\epsilon_0$~ and ~$\Omega (t)$~
fluctuate due to external noise). 

The above conclusion of the absence of the
intrinsic decoherence for the population
difference in the case of intrinsic
~$su(2)$~ symmetry is the main result of this work.
It is important 
that it is not limited by the specific form of the Hamiltonian
(\ref{1_1})-(\ref{1_4}). 
In fact, any form of ~$H$~ where ~$U_1=U_2$~ and the 
interaction part can be expressed solely in terms
of the total density operator ~$\hat{\rho}=
\Psi_1^{\dagger}\Psi_1+
\Psi_2^{\dagger}\Psi_2$~ will lead to the
same conclusion. This is so because
~$H$~ and ~$ J_{z,\pm}$~
form a closed algebra in this general case as well.

Now let us consider particular cases.
First, we study the effect
of the two short pulses investigated experimentally
in Ref.\cite{JILA}. 
In the initial state
~$|t=0\rangle$, all ~$N=N_1+N_2$~ bosons
occupy only one component (e.g.,$i=2$),
so that ~$\langle t=0|J_z|t=0 \rangle= N/2$~
\cite{JILA}. 
The
effect of a short pulse can be described by
the unitary transformation ~$U=\exp(-i\int dt H_P)$~
in the sudden approximation (from now on we
choose units ~$\hbar =1$). Taking into account
Eq.(\ref{1_4}), we find

\begin{eqnarray}
\displaystyle U={\rm e}^{-i(v'J_x + v''J_y)},
\label{5}
\end{eqnarray}
\noindent
where ~$v'=\int dt \Omega'(t),\,\, v''=
\int dt \Omega''(t)$, with ~$\Omega'(t)$~
and ~$\Omega''(t)$~ being the real and 
imaginary parts of ~$\Omega (t)$ in Eq.(\ref{1_4}). 
As a result of the pulse taking place at some
time moment $t$, any operator ~$\hat{O}(t)$~
before the pulse transforms into ~$U^{\dagger}
\hat{O}(t)U$~ after the pulse.
Specifically, the operator ~$J_z$~ changes
as 

\begin{eqnarray}
J_z \to U^{\dagger}J_zU=\cos v J_z +
{\sin v \over v}\left(v'J_y
-v'' J_x\right)\,\, ,
\label{5_1}
\end{eqnarray}
\noindent
where ~$v=\sqrt{v '^2 + v''^2}$.
This relation is a consequence
of (\ref{3_1}).
Let us choose
$v=\sqrt{v '^2 + v''^2}=\pi/2$, which corresponds
to the $\pi/2$-pulse \cite{JILA}. 
If the first ~$\pi/2$-pulse was imposed at
time $t=0$, and the ~$-\pi/2$-pulse
was produced at ~$t >0$,
the state just after the second pulse
can be represented
as ~$ |t\rangle =U^{\dagger}\exp (-iHt)U|t=0\rangle
=\exp(-iU^{\dagger}HUt)| t=0\rangle  $. 
Then, the population difference $N_2-N_1=2J_z$ after
the second pulse is

\begin{eqnarray}
\displaystyle 
2\langle t|J_z|t\rangle =
2\langle t=0|{\rm e}^{i\epsilon_0t U^{\dagger}J_zU}
J_z{\rm e}^{-i\epsilon_0t U^{\dagger}J_zU}
|t=0\rangle=
N\cos(\epsilon_0 t)\,\, .
\label{7}
\end{eqnarray}
\noindent
where the explicit form (\ref{1_1})-(\ref{1_4})
(for ~$H$~ with ~$g'=0,\,  \tilde{g}=0,\, H_{1a}=0$) 
as well as
the conditions
~$ \langle t=0|J_{x,y}|t=0\rangle =0$, and
Eq.(\ref{5_1}) have been employed
for ~$v=\pi/2$. We also took into account
that ~$[J_j, H_S]=0$. 

Eq.(\ref{7}) represents an exact
many body solution in the presence of the binary
collisions for the case of exact intrinsic ~$su(2)$~
symmetry. 
The collisions, causing dissipation of the 
normal modes, do not affect the dynamics of
the $su(2)$ generators ~$J_j$~ and their observables,
as long as the
intrinsic ~$su(2)$~ symmetry holds \cite{COM2}.
As a matter of fact, in the above
analysis we never assumed the presence of the
Bose-Einstein condensate.
Thus, the solution (\ref{7}) is formally
true in the normal phase as well.  
In some sense, the intrinsic ~$su(2)$~ symmetry
$protects$ the population difference from decoherence.
This property can be employed in quantum
computations \cite{QC}, where decoherence is usually
the main fundamental obstacle.

The goal of the present paper is 
elucidating the role of the intrinsic
~$su(2)$~ symmetry rather than 
discussing all possible mechanisms of
the decoherence, which arise due
to violation of the symmetry. 
This will be done in a separate publication.
Nevertheless, let us discuss some 
most obvious sources
of the decoherence. First,
we comment on the role of    
instrumental noise which is always 
present. Referring to the solution
(\ref{7}), the noise may produce fluctuations
of ~$\epsilon_0$~. In general, these can be 
of two types: 
i) temporal ~$\epsilon_{ir}(t)$~
with some
short correlation time ~$\tau_{ir}$~; ii) ensemble 
fluctuations ~$\epsilon_{r}$~ from realization to 
realization (in the destructive measurements
\cite{JILA,JILA2}), so that ~$\epsilon_{r}=const$~
during each realization. 
Thus, one can represent ~$\epsilon_0=
\overline{\epsilon}_0 + \epsilon_{ir}(t)
+\epsilon_{r}$~, where ~$\overline{\epsilon}_0=
\langle \epsilon_0\rangle$~
stands for the mean over the fluctuations
(which are assumed to be gaussian).
The type i) produces
irreversible decoherence characterized by 
the exponential decay ~$\langle t|J_z|t\rangle
\sim \exp(-\Gamma t)$~, with ~$\Gamma
=\int^{\infty}_0 dt \langle \epsilon_{ir}(t)
\epsilon_{ir}(0)\rangle \approx \tau_{ir}
\langle \epsilon_{ir}(0)
\epsilon_{ir}(0)\rangle $~
at ~$t\gg \tau_{ir}$.
The type ii) induces reversible decoherence (dephasing)
characterized by the gaussian time dependence
~$\langle t|J_z|t\rangle 
\sim \exp(-t^2/\tau_d^2)$~ with the
dephasing rate ~$\tau_d^{-1}=\sqrt{\langle 
\epsilon^2_{r}\rangle/2}$~.
The reversible dephasing will dominate the
evolution if the condition ~$\tau^{-1}_d \ll \Gamma^{-1}$~
is fulfilled, provided ~$\tau_d \gg \tau_{ir}$. 
In other words, the irreversible temporal noise (type i))
is irrelevant if
~$\sqrt{\langle 
\epsilon^2_{r}\rangle }\gg \tau_{ir}
\langle \epsilon^2_{ir}(0)\rangle$ .
Thus, the decay is reversible at times
~$\tau_d < t < \Gamma^{-1}$.
This can be tested
by the echo method \cite{ECHO,ECKUK,ECEXP}.
In order to observe the echo,
three pulses are required, so that the
first ~$\pi/2$-pulse takes place at $t=0$, then some 
extra (time-reversal) pulse is imposed at ~$t=\tau >0$~, and, finally,
the readout ~$\pi/2$-pulse takes place at some time ~$t> \tau$.

Below we present an exact echo solution  
in the case of the exact intrinsic 
$su(2)$ symmetry for the case
~$\tau_d < t < \Gamma^{-1}$. The time-reversal
pulse is taken as the ~$-\pi$-pulse. 
Then, the state just after the last pulse 
is ~$
 |t\rangle =U\exp(-iH(t-\tau))
U^{\dagger} U^{\dagger}\exp(-iH\tau)U|t=0\rangle=
\exp(-iUH U^{\dagger}(t-\tau))
\exp(-i U^{\dagger} HU\tau)|t=0\rangle $.
We take into account 
(\ref{3_1}),
and obtain
the population difference ~$N_2-N_1 =2\langle t|J_z|
t \rangle$~  
(for ~$\tilde{g}=0,\, g'=0,\, H_{1a}=0$~) after the
third pulse as

\begin{eqnarray}
\displaystyle N_2-N_1= 
N\cos(\epsilon_0 (t-2\tau))\, .
\label{9}
\end{eqnarray}
\noindent
The solution (\ref{9}) gives a 
complete revival of the initial
population difference at $t=2\tau$. 
Averaging
over the type ii) noise yields 

\begin{eqnarray}
N_2-N_1= N\cos(\overline{\epsilon}_0 
(t-2\tau)){\rm e}^{-(t-2\tau)^2/\tau_d^2}\, .
\label{90}
\end{eqnarray}
\noindent
Should the temporal noise (the type i)) be
present, the echo (at ~$t=2\tau$~) acquires
the suppression factor ~$\exp(-2\Gamma \tau)$~.
Thus, the echo can be observed if ~$2\tau \Gamma \leq 1$.

An intrinsic decoherence 
of ~$\langle t|J_z|t \rangle$~  
occurs when the intrinsic $su(2)$
symmetry is broken. This may be due to
deviations of ~$\tilde{g},\, g'$~ and ~$H_{1a}$~ from 
zero \cite{COM1}. Accordingly, the terms describing coupling
of the operators (\ref{3}) to the normal
modes will enter Eqs.(\ref{4}). We note that
the effect of ~$H_{1a}\neq 0$~ has been studied
recently in Ref.\cite{WILLI}. 

Here, we
consider the effect of the shot noise
produced by uncertainty in the initial depostion
of the total ~$N$~ in the trap. In this analysis we 
take into account that ~$\tilde{g}=0,\, g'\neq 0$~ 
in Eqs.(\ref{1_2}),(\ref{1_3}) for 
the case \cite{JILA}. We
also assume that the shot noise is large, so that
thermal effects can be neglected, and one can set
~$T=0$.
Furthermore,
we consider no asymmetry in the trapping potential
(~$H_{1a}=0$).
At ~$T=0$~  the main contribution
to  ~$\langle t| J_z|t \rangle= (N_2-N_1)/2$~
comes from such many-body eigenstates where almost
all ~$N_{1,2}$~ belong to the 
corresponding condensate single particle
states ~$\varphi_{01}({\bf x}),\, \varphi_{02}({\bf x})$.
In this case, 
the operators ~$\Psi_{1,2}({\bf x})$~ can be represented as
~$\Psi_1=a_1\varphi_{01}({\bf x}),\, $~
and ~$\Psi_2=a_2\varphi_{02}({\bf x})$, where
~$a_{1,2}$~ destroy one boson from the condensate
states
~$\varphi_{01}({\bf x}),\, \varphi_{02}({\bf x})$,
respectively. Such an ansatz resembles
the two-mode approximation widely
used previously (see, e.g., in Ref.\cite{MILBURN}).
Employing it in Eqs.(\ref{1_1})-(\ref{1_4})
and (\ref{3}) yields the Hamiltonian (\ref{1_1}) as

\begin{eqnarray}
\displaystyle 
H=C + (\epsilon_0+b_1N)J_z + b_2J^2_z + 
{\Omega^* (t)\over 2} J_+ +{\Omega (t)\over 2} J_-,
\label{16}
\end{eqnarray}
\noindent
where $C$ commutes with (\ref{3}), and
thereby can be omitted as long as the 
dynamics of ~$J_{z,\pm}$~ is concerned; 
the coefficients 
~$b_1,\, b_2$~ can be expressed in terms
of the original parameters entering 
Eqs.(\ref{1_1})-(\ref{1_3}). In fact,
the form (\ref{16}) is general as long as 
only the binary collisions are taken into
account, and no exchange interaction
occurs. It is important
that ~$b_1,\, b_2$~ vanish if
the intrinsic ~$su(2)$~ symmetry holds. 
The effect of the shot noise is described
by the term ~$\sim b_1$. In this paper
we consider the simplest case of the 
confinement in a box of the
volume ~$V$, and neglect the
phase separation effect \cite{JILA}. Then,
provided ~$H_{1a}=0,\, \tilde{g}=0$, we find
~$b_1=-2g'/V,\,\, b_2=0$~\cite{COM}
 in Eq.(\ref{16}).  
Now it can be seen that the role of
~$\epsilon_0$~ is played by ~$\epsilon_0 -2g'N/V$~
in the solutions (\ref{7}), (\ref{9}),
provided ~$N=const$~ during each measurement
\cite{JILA}. 
We assume that
the shot 
fluctuations are characterized by some known
variance $\Delta N$. This 
yields the dephasing rate

 \begin{eqnarray}
\tau_d^{-1}={\sqrt{2}|g'|\Delta N\over V}
\label{17}
\end{eqnarray}
\noindent
in Eq.(\ref{90}).
It is important to emphasize that the shot
noise does not disrupt the phase
coherence in each realization. Therefore,
the echo effect described by Eq.(\ref{90})
will be observed in this case. It is worth also noting
that the trap losses may introduce some
effective temporal noise into Eq.(\ref{16}).
It will yield the exponential factor ~$\sim \exp (-t/\tau_L)$~
in Eq.(\ref{90}), where ~$\tau_L$~ is the confinement time.
Thus, the trap losses do not
disrupt the coherence because in any realistic
experiment its duration $t$ should be much shorter
than ~$\tau_L$. 

Let us estimate the rate of the dephasing 
induced by the shot noise. In $^{87}$Rb, 
~$|g'|/g=0.03$; the quantity ~$\mu\approx gN/V$, which
is a typical
value of the chemical potential, can be 
taken as ~$\mu \approx 10^2-10^3$s$^{-1}$ 
(~$\hbar =1$) \cite{REV}.
Thus, ~$\tau_d^{-1}\approx 0.04\mu \Delta N/N
\approx (4 - 40) \Delta N/N$s$^{-1}$. 
For ~$\Delta N\approx \sqrt{N}$, we estimate
~$\tau_d\approx (0.3 -3) $s from Eq.(\ref{17}),
where we take ~$N=10^4$.  For ~$N=10^6$, this
estimate increases by the factor of 10.
Such long decoherence times should be contrasted with the
longest relaxation times $10-50$ms for the normal   
modes \cite{JILA2}. We note, however,
that the effect of the shot noise may be deliberately
enhanced by making, e.g., $\Delta N\sim N$. In this case
the echo effect can be made much more pronounced.
For example, if ~$\Delta N=0.1 N$~ for typical number
of atoms ~$N\approx 10^4$~ and 
~$\mu \approx 10^3$s$^{-1}$, we find
~$\tau_d\approx 30$ms. 

The case
of ~$T\neq 0$~ when the intrinsic $su(2)$ symmetry 
is slightly broken requires
special consideration. 
This will be presented
in a separate work. Here we only give a qualitative
reason as to why the {\it reversible dephasing} should
dominate the evolution of ~$N_2-N_1$~ between the rf-pulses
as long as
a significant portion of atoms forms the two-component
condensate. Indeed, if no external bath
is present and ~$N_{1,2}$~ are conserved separately,
 the dominant evolution 
of the condensate
operators ~$\Psi_i({\bf x}, t)\approx a_i(t)
\varphi_{0i}({\bf x})$~ 
in any given exact many-body
eigenstate 
is simply given by 
the exponents ~$\exp(-\mu_i t)$~, where
~$\mu_i$~ is the chemical potential of
the ~$i$th component in the
given eigenstate. 
A detailed analysis of this has been given
in the case of a single component condensate
in Ref.\cite{OC_1}. Thus, temperature
is simply producing the thermal averaging of
the exponent ~$\exp (i(\mu_1-\mu_2)t)$~
over the exact eigenenergies. This produces
the gaussian decay determined by the thermal
fluctuations of the relative
chemical potential ~$\mu= \mu_1-\mu_2$.        
In other words, the 
normal modes of the 
condensate do not produce irreversible
decay of the global phase \cite{COM3}. 
This is in close analogy with the case of the
single condensate \cite{OC_1}. 
Thus, the echo effect should persist
at ~$T\neq 0$~ as well, if BEC is present \cite{DAMOP}. 

In summary, we have pointed out that
the dynamics of the population difference
in the two-component atomic vapor is controlled
by the closeness to the $su(2)$ internal symmetry.
In the case of the exact symmetry, no intrinsic 
decoherence occurs for the population difference.
This result answers the question posed by
the JILA experiment \cite{JILA}.  
In the presence of the
condensate, slight deviations from the 
intrinsic symmetry
result in reversible dephasing  
of the population difference
which  can be revealed in the
echo experiment. We suggest considering
the global relative phase of the two-component
condensate, which obeys the intrinsic ~$su(2)$~ symmetry
and thereby is protected from the intrinsic decoherence,
as a viable candidate to be employed 
for quantum computations.

\acknowledgements
One of the authors (A.K.) is grateful to 
E. A. Cornell for useful discussion 
of possibilities of observing
the echo effect. This work was partially
supported by a CUNY Collaborative Grant.


\begin{references}
\bibitem{COM1}
There are certain
deviations from the perfect symmetry
in ~$^{87}$Rb.
The scattering lengths are in the ratio
~$a_1:a_{12}:a_2=1.03:1:0.97\neq 1$~
\cite{JILA2}. Eq.(\ref{2}), then, yields
~$ g'=0.03g $~ and still ~$\tilde{g}=0$. 
Furthermore, ~$U_1$~ and ~$U_2$~
are slightly different.
\bibitem{COM2}
Most of the local operators ~$\hat{O}$~
which do not commute with ~$H_S$ should
exhibit dissipation induced by the collisions
due to the term $\sim g$ in (\ref{1_2}).
\bibitem{COM}
Including
the effect of ~$\tilde{g}\neq 0$~ as well as of the
phase separation results in ~$b_2\neq 0$~ 
in Eq.(\ref{16}). This term is responsible 
for the quantum phase diffusion effect (for the
relative global phase) 
considered in
Ref.\cite{MILBURN}. If the variance of $N$ due
to the shot noise is much larger than the variance of $M$
\cite{MILBURN}, the term ~$\sim b_2$~ can safely be ignored.
\bibitem{COM3}
If some infinite external bath contacts a finite condensate,
irreversible phase decoherence should
take place \cite{LEG,SOL,OC_1}.
\end{references}
\end{document}